# Synthesis and Characterization of Compositionally Complex $(Gd/Ho/Er/Dy)_2Zr_2O_7$ Thin Film Combinatorial Library

Dalton A. Pearl[1], Jade Holliman Jr[1], Reece Emory[1], Joshua Safin[1], Aditya Raghavan[1], Kamyar Barakati[1], Andrew H. Jones[2], Ethan A. Scott [3], Jack C. Lasseter[4], Adam Corrao[5], Daniel Olds[5], Bruce Ravel[6], Sergei K. Kalinin[1], Patrick E. Hopkins[2,3,7,8], Katharine Page[1], Philip D. Rack[1*]

1. The University of Tennessee, Knoxville, Department of Materials Science and Engineering, Knoxville TN 37996, USA
2. Laser Thermal, Charlottesville VA 22902, USA
3. The University of Virginia, Department of Mechanical and Aerospace Engineering, Charlottesville VA 22903, USA
4. Center for Nanophase Materials, Oak Ridge National Laboratory, Oak Ridge TN 37830, USA
5. National Synchrotron Light Source II, Brookhaven National Laboratory, Upton NY 11973, USA
6. National Institute of Standards and Technology, Gaithersburg MD 20899, USA ORCID 0000-0002-4126-872X
7. University of Virginia, Department of Materials Science and Engineering, Charlottesville VA 22903, USA
8. University of Virginia, Department of Physics, Charlottesville VA 22903, USA

*corresponding author: prack@utk.edu

[1] Any mention of commercial products within NIST web pages is for information only; it does not imply recommendation or endorsement by NIST.

**Abstract:** High-throughput synthesis and characterization of novel ceramic materials with improved thermomechanical properties and phase stability are needed to accelerate the discovery of next-generation thermal barrier materials. A combinatorial thin film material library of $(GdDyHoEr)_2Zr_2O_7$ were created via combinatorial magnetron reactive sputtering with rare-earth/zirconium alloy targets. Structural, chemical, and thermal property characterization mapping across the four component composition space was performed and correlated with thermal transport measurements. X-ray diffraction confirms the formation of a single-phase defect fluorite structure, while energy-dispersive X-ray spectroscopy verifies a rare-earth-to-zirconium ratio of 1:1 across the library. X-ray absorption near edge structure and extended X-ray absorption fine structure measurements reveal minimal variation in cation valence states, local coordination environments, and nearest-neighbor bond distances, consistent with the EXAFS confirm the defect fluorite structure and trends in the associated ionic radii. Steady state thermoreflectance mapping identifies a pronounced minimum in thermal conductivity within the Dy/Gd-rich quadrant. This minimum does not coincide with either the equiatomic composition or the region predicted to exhibit maximum cation size disorder. Instead, it corresponds to the largest experimentally observed lattice parameter, despite deviating from Vegard-like chemical averaging, and is independent of grain size and whole-pattern microstrain. These observations suggest that the way the fluorite lattice accommodates compositional complexity, rather than cation size disorder alone, provides a more informative descriptor of thermal transport. Overall, this work establishes a high-throughput workflow for combinatorial thin-film synthesis and multimodal characterization, enabling the rapid identification of previously inaccessible structure-property relationships in compositionally complex ceramics.

Keywords: Combinatorial Magnetron Reactive Sputtering, Thin films, Compositionally Complex Oxides, Thermal Conductivity

[1] Any mention of commercial products within NIST web pages is for information only; it does not imply recommendation or endorsement by NIST.

## 1. Introduction

Thermal barrier coatings (TBCs) typically comprise a ceramic topcoat layer that protects and insulates, for instance, the components of advanced combustion engines, such as those used in aircraft and power generators.[1], [2], [3] The use of ceramic TBCs can reduce surface temperatures by several hundred degrees, thereby prolonging component lifetime and improving engine efficiency. Currently, yttria-stabilized zirconia (YSZ) is the commercial standard for thermal barrier coatings (TBCs) due to its low and constant thermal conductivity over a large temperature range (~ YSZ (7 mol% $Y_2O_3$), 2.3 $W(m\ K)^{-1}$), high tolerance of thermal shock, and oxidation resistance.[4] However, after prolonged high-temperature cycling, phase instability and sintering effects ultimately lead to failure of the TBC.[5] Specifically, the metastable t′ phase in 7YSZ becomes increasingly unstable above ~1200-1300 °C, limiting its use in higher-temperature applications. Thus, next-generation turbine systems with higher combustion chamber temperatures and pressures require novel ceramic materials with lower thermal conductivities and high temperature phase stability.

Rare-earth zirconates (REZrOx), of the $RE_2Zr_2O_7$ type, have been explored for next-generation TBCs due to their lower thermal conductivities and higher temperature phase stabilities compared to YSZ. REZrOx crystallizes into the ordered pyrochlore phase if the ionic radius ratio between rare-earth and Zr atoms falls between 1.46-1.78 (La through Dy), however, it favors the disordered defect fluorite phase for smaller values (Ho through Lu, and Y).[6], [7] Likewise, at higher temperatures/pressures, the defect fluorite phase is observed.[8], [9] Wu *et al*[10] explored the thermal conductivities of several REZrOx fluorite-phase ceramics. The thermal conductivities of pyrochlore $Gd_2Zr_2O_7$, $Nd_2Zr_2O_7$, and $Sm_2Zr_2O_7$ reported are 1.6, 1.6, and 1.5 $W\ (m\ K)^{-1}$ respectively. Synthesis and processing variations can induce metastable / tunable structure variations, particularly near the Dy/Ho phase boundary. Furthermore, Vassen *et al.*[11] found that a pyrochlore $La_2Zr_2O_7$ ceramic has a thermal conductivity of 1.6 $W\ (m\ K)^{-1}$ at 1000 °C. Although the thermal expansion coefficient (TEC) of $RE_2Zr_2O_7$, which must match that of the underlying substrate to minimize thermal stress, requires further enhancement to meet this standard, as well as lifetime durability.[12]

High-entropy oxides (HEOs), which form a single-phase ceramic with multiple cations, have garnered interest due to their lower thermal conductivity, enhanced mechanical properties, corrosion resistance, and improved phase stability compared to single-component systems.[13] The improved properties of HEOs are attributed to 4 main factors: (i) lattice distortion due to size disorder amongst lattice sites, (ii) high entropy phase stabilization, (iii) slow diffusion, and (iv) cocktail effects linked to mixing of principal elements.[13], [14] Compositionally complex materials  High entropy oxides are qualitatively characterized based on the calculated configurational entropy ($S_{config}$) of the crystal system. Compositionally complex materials are more general class of materials where $S_{config}<1.5R$ (R being the gas constant) but still incorporate multiple cations on a single lattice position.[15] Compositionally complex rare-earth zirconates (CCREZrOx) exploit these effects through substitution of the zirconate system with multiple rare-earth elements or multiple transition metal elements or combinations thereof. Liu *et al.*[16] synthesized an equiatomic five rare-earth (RE=La, Nd, Sm, Gd, Yb) CCREZrOx via high-

[1] Any mention of commercial products within NIST web pages is for information only; it does not imply recommendation or endorsement by NIST.

speed positive grinding and solid-state reaction. The system exhibited a single-phase defect fluorite crystal structure and a low thermal conductivity of 0.9–1.72 W (m K)^-1 at temperatures ranging from 273 to 1273 K. Similarly, Zhao *et al.*[17] reported a 5 component REZrOx pyrochlore of La, Ce, Nd, Sm, and Eu with an amorphous-like ultra-low thermal conductivity of 0.76 W*(m*K)^-1 at room temperature, which exhibited decreased grain growth compared to the single-component counterpart LaZrOx. Ren *et al*[18] published work on a $(Sm_{0.2}Eu_{0.2}Tb_{0.2}Dy_{0.2}Lu_{0.2})_2Zr_2O_7$ defect fluorite, with a similar thermal conductivity of 0.86 W $(m K)^{-1}$ at 1273 K. Luo *et al*[19] correlated the CCREZrOx space to various thermophysical properties such as electronegativity difference, atomic mass difference, and ionic radius difference. It was found that the TEC was negatively correlated with the electronegativity difference and lower thermal conductivity, both of which increased with increasing atomic mass difference and ionic size difference. This aligns with the findings of Wright *et* al[20], where 18 medium to high-entropy pyrochlore ceramics were synthesized via high-speed ball milling. In this study, researchers found that increasing the size disorder, as measured by the standard deviation of ionic radius in the sublattice (in the pyrochlore structure, there are two cation sublattices at the A and B sites), correlated with a lower thermal conductivity, suggesting the importance of severe lattice distortion for low thermal conductivity. Current research on CCREZrOx focuses on materials synthesized via bulk synthesis routes, with only a handful of published works using alternative synthesis routes, such as atmospheric plasma spraying[21], electron-beam physical vapor deposition (EB-PVD)[22], [23], sol-gel, and electrospinning.[24], [25]

With the expansiveness of the HEO material space, a higher-throughput methodology is needed to explore material systems and correlate structure and properties. Reactive magnetron sputtering (RMS) is a potential solution for rapid material synthesis of thin films. RMS allows for precise control over microstructure, chemical composition, and structural properties.[26], [27], [28] Lin *et al*[29] reported a multi-component (AlCrTaTiZr)Ox thin film that was amorphous as-deposited, while a post-anneal above 900 °C precipitated several oxide phases. Kirnbauer *et al.*[30] synthesized a five-cation HEO via RMS using a single equimolar Al-Cr-Nb-Ta-Ti target. Employing a substrate temperature of 400 °C, the as-deposited samples demonstrated a rutile crystalline structure. The material system also demonstrated high-temperature phase stability (1200 °C) with a preferred orientation in the [101] direction and improved mechanical hardness compared to its ternary oxide sister compounds. Furthermore, to the best of our knowledge, no work has been published to date exploring non-equiatomic cation concentrations in $CCRE_2Zr_2O_7$. Thus, we explore a $CCRE_2Zr_2O_7$ thin film library synthesized via combinatorial reactive magnetron sputtering (CRMS) with rare-earth cation concentration gradients.

In this work, the thermal conductivity of a $(GdDyHoEr)_2Zr_2O_7$ CCREZrOx, library synthesized via CRMS of four rare-earth-zirconium alloy targets in an Argon/$O_2$ atmosphere, is explored. The thin film's structural properties were investigated using both laboratory and synchrotron X-ray diffraction . Chemical gradients, microstructure, and surface topography were measured via energy-dispersive X-ray spectroscopy, X-ray diffraction, and atomic force microscopy, respectively. Local bonding environments were determined via X-ray absorption near edge structure (XANES) and X-ray absorption fine structure (XAFS). Thermal conductivity was mapped via steady-state thermoreflectance (SSTR) and correlated to the structural, chemical, and physical property gradients. This work demonstrates the correlation between various physical properties (size disorder and configurational entropy) and low thermal conductivity in CCOs. Furthermore, the study lays the groundwork for a high-throughput materials discovery and analysis methodology that can be utilized in future work to identify non-equiatomic material compositions with specific properties, thereby expediting the discovery of compositionally complex alloy and ceramic materials.

[1] Any mention of commercial products within NIST web pages is for information only; it does not imply recommendation or endorsement by NIST.

## 2. Experimental Methods

### 2.1 Combinatorial Reactive Magnetron Sputtering

The $(GdDyHoEr)_2Zr_2O_7$ thin films were deposited at room temperature by combinatorial reactive magnetron sputtering in an argon and oxygen atmosphere. The films were co-sputter deposited using an AJA 2000 Sputtering system (AJA International Inc) from alloy targets of Gd/Zr, Dy/Zr, Ho/Zr, and Er/Zr (50/50 at%). Targets were made via vacuum arc melting (Kurt J. Lesker Company) with 99.9% purity and dimensions of 2.00" diameter and 0.25" thickness. Films were deposited onto 100 mm diameter fused silica and Si. The four targets are arranged in a sputter-up geometry where the substrate occupies the center of a projected square and the four targets are located on the square corners and tilted at an angle of 37° relative to the substrate. For a complete system overview see *Fowlkes et al.*[31]

Before deposition, the vacuum chamber was pumped to a pressure of $5 \times 10^{-7}$ Torr ($6.7 \times 10^{-5}$ Pa). The flow rate of argon/oxygen was 25/4 SCCM and metered with a pressure control valve to maintain a 5 mTorr (0.67 Pa) processing pressure. The argon gas was injected at the targets, and oxygen at the substrate to enhance the substrate oxidation and extend the metal-mode reactive sputtering regime. The applied power to each target alloy was 200 W RF for the Ho/Zr and Er/Zr, while Dy/Zr and Gd/Zr were powered with 200 W DC to approximately achieve equal sputtering rates from each target. A 10 W DC substrate bias was applied to reduce the porosity in the as-deposited films [see Supplemental Information Figure SI4 for a Scanning Electron Microscopy (SEM) image for a thin film made without substrate bias]. Films were sputtered for 40 minutes to achieve a center thickness of 1.3 µm.

### 2.2 Chemical and Structural Characterization

#### Energy Dispersive X-ray Spectroscopy and Scanning Electron Microscopy

Energy-dispersive X-ray spectroscopy (EDS) was used to measure the composition at various positions across the combinatorial library and Scanning Electron Microscopy (SEM) was used to measure cross-section thickness at a single location. EDS and SEM were performed on a Helios PFIB dual beam system, with an acceleration voltage of 20 kV, and an Oxford Instruments Ultim Max Infinity EDS detector integrated on the instrument. The sample prepared for EDS/SEM was deposited on a silicon wafer to minimize charging.

#### X-ray Diffraction

---

[1] Any mention of commercial products within NIST web pages is for information only; it does not imply recommendation or endorsement by NIST.

For initial phase identification, X-ray diffraction (XRD) of the films was performed using a PANanalytical MRD diffractometer with an optical component configuration ¼ divergence slit, 4 mm mask, 0.04 radians Soller slit, 10 to 80 2θ scan range, 0.1° step rate, 3° omega offset, and 2θ-ω scan axis. Cu Kα radiation of $\lambda = 1.54$Å was used for the incident X-ray. A Ni filter was used to reduce Kβ components. The obtained data were analyzed using GSAS-II software.[32]

Synchrotron XRD

Synchrotron X-ray diffraction data were collected in Debye-Scherrer (transmission) geometry on the Pair-Distribution Function (PDF) beamline 28-ID-1 at the National Synchrotron Light Source-II (NSLS-II). Measurements were made using an incident beam energy of 74.46 keV ($\lambda = 0.1665$ Å) with a 0.4 mm x 0.4 mm beam size and were recorded on a flat-plate CsI area detector (PerkinElmer XRD 1621, 0.2 mm x 0.2 mm pixel size, 2048 x 2048 pixel array). Conventional diffraction measurements were completed using a sample-to-detector distance of 997.838 mm, providing data out to $Q_{max} = 10.40$ Å$^{-1}$ ($d_{min} = 0.6042$ Å). The total measurement time was 30s, consisting of 30 individual 1s exposures summed on the detector.

Instrument geometry (sample-to-detector distance, detector tilts) was determined using the pyFAI[33], [34] calibration tool on 2D diffraction images collected on a standard reference materials (NIST SRM 660c, $LaB_6$, $a$=4.15682(8) Å). Detector image masks were constructed to exclude the beam stop, detector edges, and dead pixels. Azimuthal integration of images to conventional 1D patterns was then done using the standard data processing routines at the beamline. Pattern intensities were then normalized and background subtracted by summing the intensity over a small 2θ (°) range where only the amorphous $SiO_2$ background signal was present, rescaling to the background signal, and subtracting the background.

The instrumental contribution to the diffraction peak profiles was determined by fitting a Thompson-Cox-Hastings pseudo-Voigt function[35] to data collected on a standard reference material (NIST SRM 660c, $LaB_6$, $a$=4.15682(8) Å). Additionally, a 3-term polynomial was refined to correct for a minor 2θ-dependent peak position offset due to detector parallax effects. The instrument profile function and peak position correction terms were then fixed in subsequent analysis of data collected on the thin films, enabling the deconvolution of instrument- and sample-based peak broadening.

Whole pattern fitting (WPF) of the diffraction data was done using the Pawley[36] method in which peak positions are modeled by a unit cell, while the peak intensities are freely refined without reference to a structural model. This enabled accurate determination of lattice parameters and peak broadening effects without bias from a structural model. Refinements were done using both a $Fm\bar{3}m$ (space group #225) and $Fd\bar{3}m$ (space group #227) unit cell model. While the $Fd\bar{3}m$ model appears to lead to better fit quality, this is simply due to the larger number of peaks present that can compensate for small misfits due to $hkl$-dependent peak shifting (*e.g.,* larger $d$-spacing for all $h$00 reflections relative to those predicted by WPF). The appropriate model for peak broadening was determined through trial fits using the Double-Voigt[37], [38] approach in which broadening due to crystallite size (cosθ dependent) and lattice strain (tanθ dependent) have separate Lorentzian and Gaussian components. It was found that strain broadening dominates, and we are not sensitive to crystallite size effects. While the Lorentzian strain contribution is largest, a better fit is achieved by including a Gaussian strain component. The size-strain behavior was further confirmed by Williamson-Hall[39] analysis in which peak breadths (β) are plotted as a function of the scattering vector ($s = 1 / d$), and size and strain are determined from the *y*-intercept (1 / intercept = volume-weighted mean column length) and slope ($\varepsilon = \Delta d/d$), respectively. All synchrotron XRD refinements were completed using TOPAS[40] (version 7, Bruker-AXS).

[1] Any mention of commercial products within NIST web pages is for information only; it does not imply recommendation or endorsement by NIST.

X-ray Absorption Fine-Structure Spectroscopy (XAFS)

XAFS measurements were carried out at beamline 6-BM Beamline for Materials Measurement (BMM) of the National Synchrotron Light Source II (NSLS-II) located at Brookhaven National Laboratory (Upton, NY, USA). Experiments were conducted at room temperature in transmission mode with ionization chambers filled with optimal mixtures of He/$N_2$/Ar/Kr gas. The energy of the incident beam was tuned with a Si(111) monochromator in the energy range 4.0 keV – 21 keV. Fluorescence measurements were collected simultaneously with a four-element vortex Si-drift detector. Fluorescence data were analyzed for all edges in the combinatorial library due to poor signal-to-noise ratio in transmission. Transmission spectra were analyzed for all edges in the bulk powder samples. XANES and EXAFS spectra were collected at the Zr K-edge (17998 eV), Er $L_2$-edge (9264 eV), Dy $L_2$-edge (8581 eV), Ho $L_3$-edge (8071 eV), and Gd $L_3$-edge (7243 eV). Er and Dy $L_2$-edges were chosen over their $L_3$-edges because the nearby Gd $L_1$- and $L_2$-edges overlap with the Er and Dy $L_3$ post-edge regions, respectively. XANES and EXAFS spectra were processed using *Athena* and *Artemis* from the *Demeter* program suite[41]. After background subtraction and normalization of the absorption spectra, the EXAFS signals were converted to wavenumber (k) space, weighted by $k^2$ to better resolve high-k signals, and Fourier transformed (FT) to pseudoradial (R) space. The Zr K-edge FT exhibited a peak around 1 Å that was attributed to multi-electron excitation (MEE), so it was subtracted for analysis of the bond distances using the MEE removal tool in *Athena*. This is a known effect in transition metal K-edge and lanthanide L-edge spectra[42], [43], [44], [45], [46]. MEE removal was attempted for the RE edges, however, it did not result in complete removal of the peak, so this step was omitted for RE edge data. Quantitative analysis of the nearest neighbor (NN) oxygen for each cation was performed by fitting the FT EXAFS signals using the *Artemis* software and FEFF6[47]. Fitting parameters include absorption edge $E_0$, RE/Zr-O distances, and mean square relative displacements (MSRD) (i.e., thermal and static motion) while amplitude reduction factor $S_0^2$ was set to 1.0 and degeneracy N (i.e., coordination number) was fixed to 7 to model the defect-fluorite structure.

Due to the time-consuming nature of XAFS measurements (~7 min / absorption edge) and large number of edges to measure, a subset of 8 positions was selected using K-means clustering [48], [49], [50], [51], [52] of the XRD measurements. K-means clustering is an algorithm that groups data into *N* clusters by repeatedly calculating the Euclidean distance between each data and the average location of data within the cluster (*i.e.,* the centroid or cluster center) and assigning each data to the cluster whose centroid is closest until the clusters no longer change. We used the implementation of this algorithm available in scikit-learnto reduce each 1D diffraction pattern from a vector with thousands of elements to an integer label based on which cluster each diffraction pattern was assigned to. From the coordinate maps colored by the labels assigned from K-means on 1D patterns we know that our data varies smoothly in space because the labels are consistent within each region (i.e., we don't see speckled maps). We completed single peak fitting on the first 3 diffraction peaks to extract their intensity, position, and full width at half maximum, and used those values as input to K-means clustering. The advantage of clustering based on the extracted peak parameters rather than the full array of intensities  is that it minimizes contributions from extraneous features such as the substrate background and air scatter.

[1] Any mention of commercial products within NIST web pages is for information only; it does not imply recommendation or endorsement by NIST.

Automated Atomic Force Microscopy (A-AFM)

Automated atomic force microscopy (A-AFM) was combined with multi-objective Bayesian optimization to enable data-efficient exploration of nanoscale surface morphology under realistic experimental time constraints. All measurements were performed using the topographic height channel over a 2 µm × 2 µm field of view discretized into 256 × 256 pixels, corresponding to a lateral resolution of approximately 7.8 nm per pixel. With a scan rate of 1 Hz, each AFM acquisition required approximately 4.15 minutes, making exhaustive sampling of the experimental parameter space impractical and motivating the use of adaptive optimization.

From each AFM image, two complementary morphological descriptors were extracted to quantify both local and mesoscale structure. Particle size was characterized using a Voronoi-based analysis of local maxima in the height map. Peak positions $\{\mathbf{r}_i\}$ were identified using a height and separation threshold, and a Voronoi tessellation was constructed over the scanned area. For each particle $i$, the area of the associated Voronoi cell $\boldsymbol{A_i}$ provides an estimate of the local particle footprint, from which an effective particle diameter was defined as

$$d_i = 2\sqrt{\frac{\boldsymbol{A_i}}{\pi}} \qquad [1]$$

The mean particle diameter,

$$\overline{d} = \frac{1}{N}\sum_{i=1}^{N} d_i \qquad [2]$$

was used as the first objective. This metric captures average particle size while remaining robust to spatial variations in particle packing.

Longer-range morphological organization was quantified through the height–height autocorrelation function,

[1] Any mention of commercial products within NIST web pages is for information only; it does not imply recommendation or endorsement by NIST.

$$C(r) = \langle h(x)h(x+r)\rangle x \qquad [3]$$

which was radially averaged to obtain a one-dimensional correlation function. The structural correlation length $\boldsymbol{\xi}$ was extracted by fitting an exponential decay,

$$\mathrm{C(r)} \sim exp(-\frac{r}{\xi}) \qquad [4]$$

over intermediate length scales. The correlation length serves as a quantitative measure of mesoscale ordering and surface uniformity and was used as the second optimization objective.

The two objectives were treated jointly within a multi-objective Bayesian optimization (MOBO) framework. The experimental control variables were represented by a vector **x**, and the objectives were combined into

$$f(x) = \begin{bmatrix} f_1(x) \\ f_2(x) \end{bmatrix} = \begin{bmatrix} \overline{d}(\mathrm{x}) \\ \xi(\mathrm{x}) \end{bmatrix} \qquad [5]$$

each objective was modeled independently using a Gaussian process ($GP$) surrogate,

$$f_k(x) \sim \mathcal{GP}(\mu_k(x), K_k(x, x')) \qquad [5]$$

with squared-exponential covariance kernels. Objective values were normalized to zero mean and unit variance to ensure balanced learning and acquisition across objectives.

The optimization was initialized with six randomly selected experiments, after which twenty additional experiments were selected sequentially. At each iteration, the next experiment was chosen by maximizing a multi-objective expected improvement ($EI$) acquisition function. Expected improvement prioritizes experimental conditions that are predicted to improve upon the current set of non-dominated solutions while accounting for predictive uncertainty in the GP models. In this context,

$$EI(x) = E[max(f^* - f(x), 0)] \qquad [6]$$

where $f^*$ denotes the current Pareto-optimal objective values. This formulation naturally balances exploration of uncertain regions of the parameter space with exploitation of regions expected to yield improved morphological outcomes.

The impact of this strategy is evident in the evolution of the Pareto front in the objective space defined by $(\bar{d}, \xi)$. During early optimization steps, the Pareto front comprised a small number of widely spaced points, reflecting sparse sampling and high uncertainty in the surrogate models. As additional AFM scans were acquired, the Pareto front expanded and became increasingly dense, revealing a well-resolved trade-off between particle size and structural correlation length. In this system, the two objectives evolved in a correlated manner, reflecting their common physical origin. Consequently, improvements in one objective were often accompanied by systematic changes in the other, enabling direct interpretation of local structure–property relationships.

---

[1] Any mention of commercial products within NIST web pages is for information only; it does not imply recommendation or endorsement by NIST.

Such correlated evolution is not universal. In many systems, objectives may decouple or directly compete, producing qualitatively different Pareto front structures. Adaptive sampling along the Pareto front therefore serves not only to identify optimal trade-offs efficiently, but also to diagnose the underlying relationships between objectives. In this way, multi-objective Bayesian optimization provides direct insight into whether structural descriptors arise from shared physical mechanisms or reflect competing structure–property pathways.

### 2.3 Thermal Property Analysis

The thermal conductivity mapping of the CCRE2Zr2O7 material library was performed via steady-state thermoreflectance (SSTR). Details of this approach are outlined by Braun *et al.* We first deposit an 80-nm aluminum film onto the sample surface by electron-beam evaporation to serve as an optothermal transducer.[53] For these measurements, we utilize the SSTR-F fiber-based thermal conductivity measurement system from Laser Thermal, which is capable of automated spatial mapping of thermal conductivity.[54], [55] The continuous wave pump and probe lasers are coaxial and are focused onto the sample. The probe wavelength is 780 nm, and the pump is 640 nm and modulated at a low frequency (1 kHz), where steady state temperature gradients are induced from the pump laser. We use a lock-in amplifier to monitor the change in reflected probe magnitude which occurs at the frequency of the modulated heating event caused by the pump. The magnitude is directly related to the thermal resistance of the sample and inversely associated with the thermal conductivity of the sample. The calibration coefficient of the transducer is obtained by calibrating the slope of the input power vs temperature rise curve with a witness sample included in the same aluminum transducer deposition. For this study, an initial 17-point radial map was measured, followed by a higher resolution 100-micron/pixel thermal conductivity map of the central 5 x 5 cm area of the sample.

## 3. Results and Discussion

### 3.1 Composition, Simulation, and Characterization

The composition mapping of the $CCRE_2Zr_2O_7$ library was first analyzed by EDS at 9 positions across the sample. All the EDS measurements confirmed the expected RE:Zr ratio of 1:1 and the cation:anion ratio of 4:7, with center cation concentration of Zr 48.6 at. %, Gd 18.3 at. %, Ho 12.9 at. %, Dy 10.6 at.%, and Er 9.5 at. %. Furthermore, cross-sectional SEM reveals a dense microstructure with an edge thickness of 1.56 μm. These parameters were used to simulate the composition space using a Knudson-Surface source model, with the input parameters being the center composition and center thickness; the center thickness was extrapolated from the SEM thickness measured on a sample witness placed at the sample edge.[31] The accuracy of the simulation to experimental composition and system-specific parameters has been shown before.[56] The simulation yields composition maps for each element as well as a corresponding thickness map of the sample (Figure 1). The composition maps demonstrate the formation of a combinatorial material library where the concentration of cations is based on distance from the target sources and deposition rates from each source. The Gd composition varies from 7 at. % to 40 at. %, whereas the Dy, Er, and Ho compositions vary from 3.5 at. % to 26 at. %, 3 at. % to 24 at.

[1] Any mention of commercial products within NIST web pages is for information only; it does not imply recommendation or endorsement by NIST.

%, and 4.5 at. % to 30 at%, respectively. In principle, Zr has a uniform composition across the target, since each target has a 50-50 atomic ratio of RE to Zr. Likewise, the thickness gradient was simulated and a gradient of 1.25 µm to 1.65 µm indicates the $Gd_2Zr_2O_7$ had a slightly higher deposition rate relative to the other sources. Tables SI1-2 contain cation concentrations for the experimental EDS versus simulated cation concentrations.

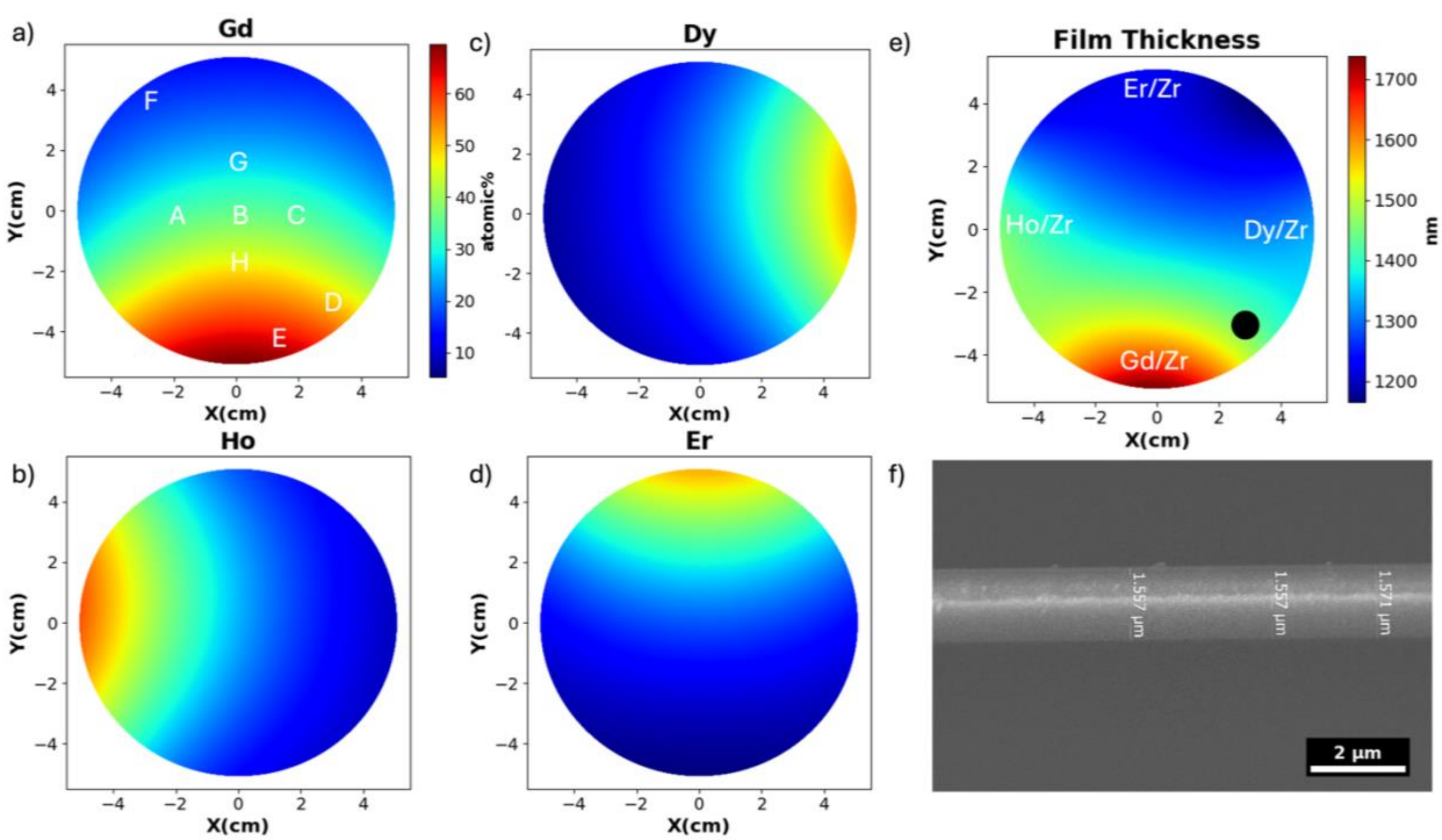

Figure 1: Simulation composition gradients of a) Gd, b) Dy, c) Ho, d) Er rare earth cations, and e) simulated thickness profile with the positions of RE/Zr targets noted. SEM image f) illustrating the dense microstructure of the film taken at the location labeled with a black circle in e). The capital letter labels A-H in panel a) indicate positions where XAFS and XANES measurements were taken.

The simulated compositions were used to calculate various properties across the compositional library, namely: average ionic radius ($\bar{r}$), size disorder ($\delta$), and configurational entropy ($S_{config}$). The $\bar{r}$ calculations follow the rule of mixtures between the rare-earth cations and Zr to determine the average ionic radius of the cation site in the $A_2B_2O_7$ defect fluorite lattice: see Table 1 for ionic radius, specified valence state, and the presumed coordination number of each element. $\delta$ is calculated from the equation

$$\delta = \sqrt{\frac{\sum_i x_i\left(1-\frac{r_i}{\bar{r}}\right)^2}{\sum_i x_i}} \qquad (1)$$

where $x_i$ is the fractional composition of the cation and $r_i$ is the ionic radius; the equation is further outlined by *Wright et al.* with a modification including the term $\sum_i x_i$ to normalize the size disorder to the cation fraction.[20] $S_{config}$ is the structural configurational entropy of both the rare earth cations and Zr and is based on the work of *Dippo & Vecchio* and is defined as, for the system, $S_{config} = -R\sum_i x_i \ln(x_i) * \frac{4}{11}$ – (2) .[15]

[1] Any mention of commercial products within NIST web pages is for information only; it does not imply recommendation or endorsement by NIST.

Table 1: Ionic radii of Gd, Dy, Ho, Er, and Zr in respective valence and coordination state.[57]

| *Element* | *Valence State* | *Coordination Number* | *Ionic Radius (pm)* |
|---|---|---|---|
| *Gd* | +3 | VIII | 105.3 |
| *Dy* | +3 | VIII | 102.7 |
| *Er* | +3 | VIII | 100.4 |
| *Ho* | +3 | VIII | 101.5 |
| *Zr* | +4 | VI | 72 |

Figure 2a-d) illustrate the resultant maps of the average ionic radius, size disorder, configurational entropy, and radius ratio of the average RE/Zr. There is a modest variation in $\bar{r}$ from 102 – 104 pm with the largest $\bar{r}$ at the Gd edge, which corresponds to Gd having the largest ionic radius. Likewise, $\delta$ is greatest along the edge nearest Gd with a value of 18.3% (the lowest value at the Ho edge is 17.1%). The largest $S_{config}$ value, 0.5R, is upshifted from the center due to the higher sputtering rate of the Gd/Zr target. As shown in d), while $Gd_2Zr_2O_7$ sits at the boundary of the defect fluorite/pyrochlore radius ratio transition, our average ionic radius ratio of RE/Zr is safely in the defect fluorite region (RE/Zr < 1.46).

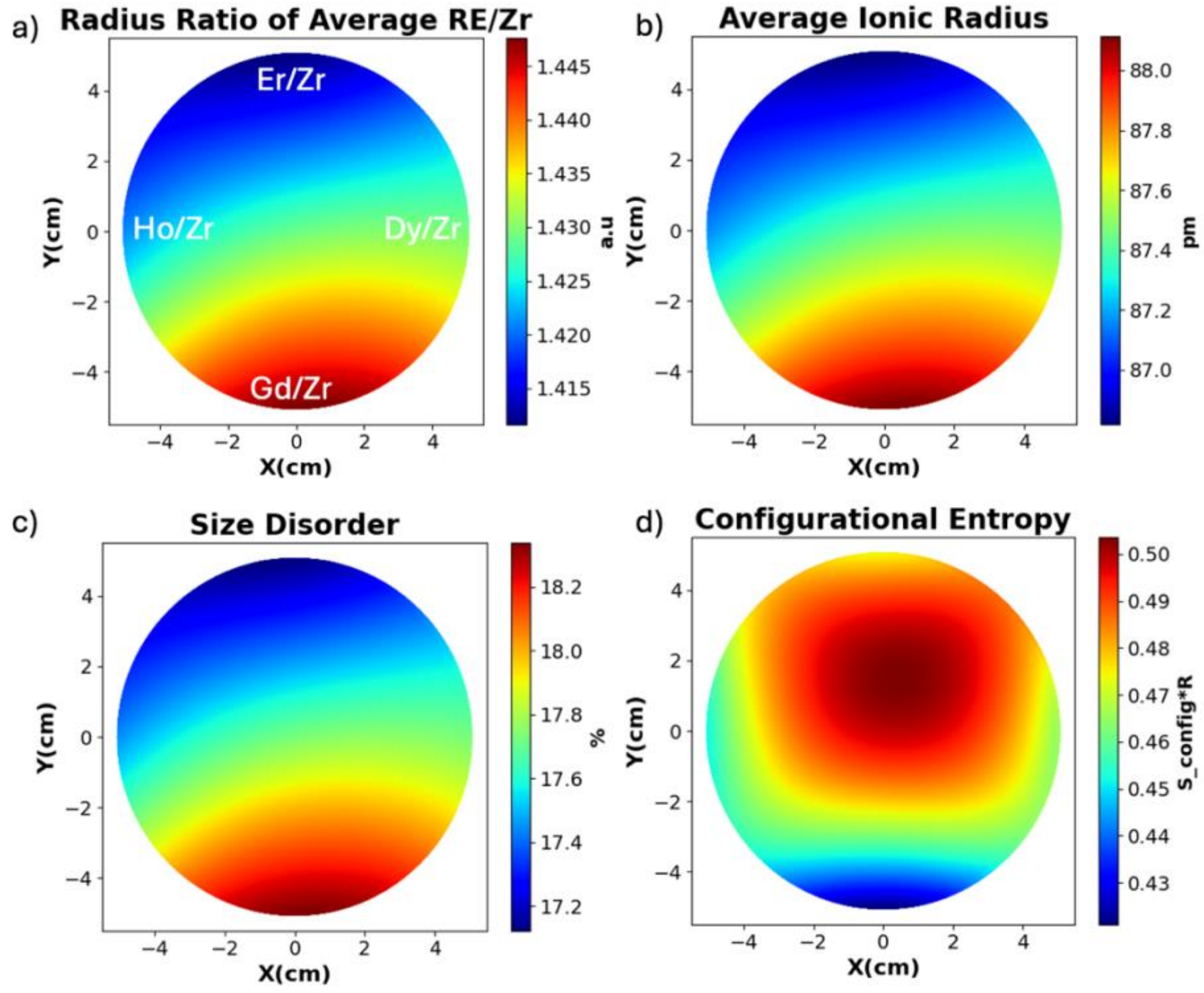

[1] Any mention of commercial products within NIST web pages is for information only; it does not imply recommendation or endorsement by NIST.

Figure 2: Simulated property gradients of a) RE/Zr radius ratio, b) average ionic radius, c) size disorder, d) and configurational entropy

To probe the chemical coordination of the films, we also performed XAFS and XANES at several locations of the film correlating to regions of interest identified by the K-means clustering of X-ray diffraction data. Figure SI2 presents the resultant K-means map, with probed locations identified. A comparison of the XANES for the Gd $L_3$- and Zr K-edges is shown in Figure 3. The XANES spectra for rare earth edges Dy $L_2$, Ho $L_3$, and Er $L_2$ are similar to the spectra for Gd $L_3$ and are included in Figure SI3.

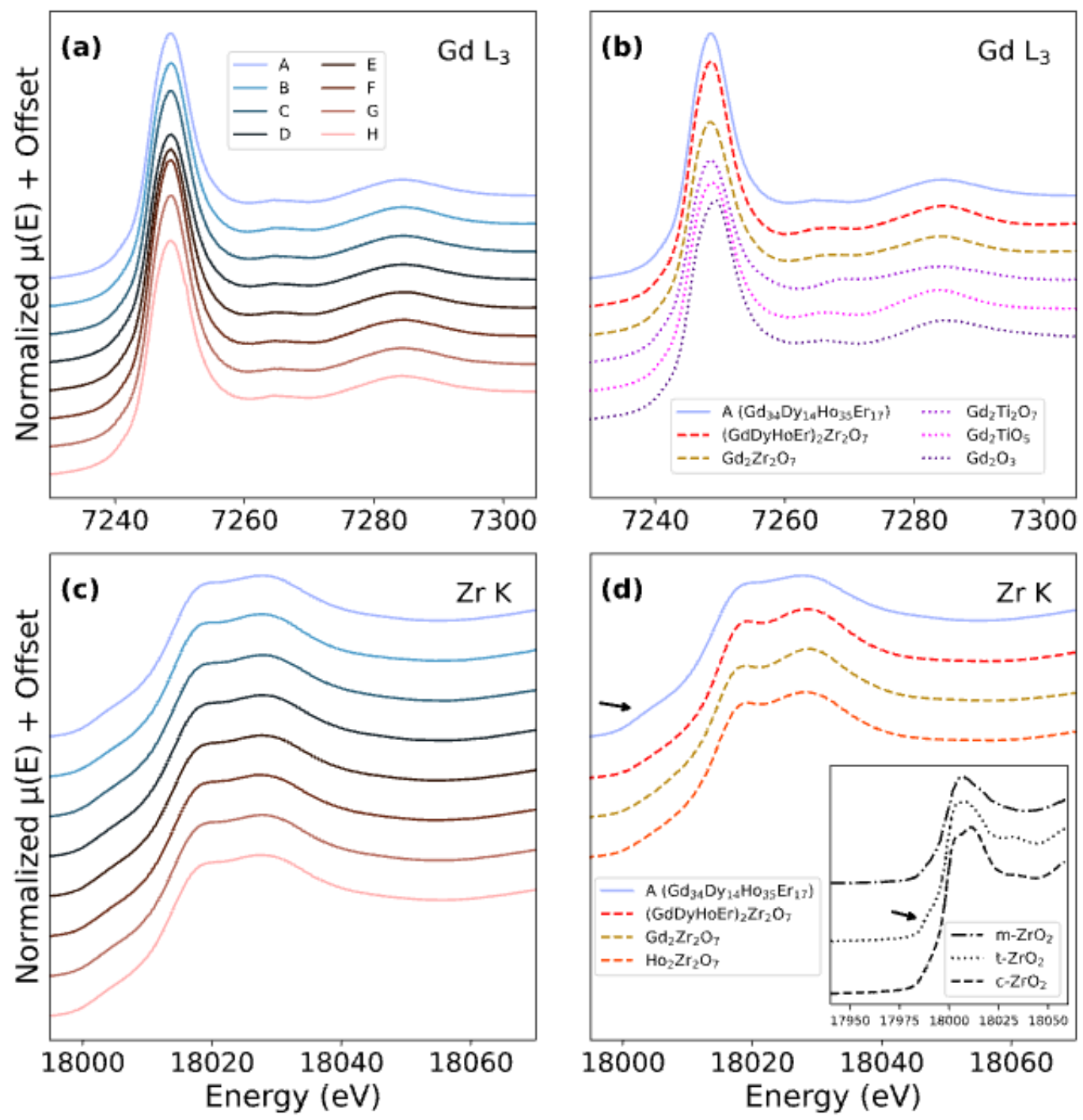

Figure 3: Normalized XANES collected for the (a) Gd $L_3$-edge and (c) Zr K-edge across $(GdDyHoEr)_2Zr_2O_x$ library positions A-H. (b) Library position A XANES compared with bulk powder compositions $(Gd_{0.25}Dy_{0.25}Ho_{0.25}Er_{0.25})_2Zr_2O_7$ and $Gd_2Zr_2O_7$ and references $Gd_2Ti_2O_7$, $Gd_2TiO_5$, and $Gd_2O_3$ for the Gd $L_3$-edge. (d) Library position A XANES compared with $ZrO_2$ polymorphs c-$ZrO_2$, t-$ZrO_2$, m-$ZrO_2$ for the Zr K-edge. The spectra are plotted with an offset for clarity. The Gd $L_3$-edge references are from the "Common XAFS Materials at BMM" database. The Zr K-edge references are from *Li* et al. [56]

The Gd and Zr edge spectra are nearly identical for positions A-H, suggesting a very uniform local structure across the library, despite the varying composition. The spectrum for position A, having a RE composition of $Gd_{34}Dy_{14}Ho_{35}Er_{17}$, is hereafter referred to as the library spectrum. There are no edge shifts indicative of valence changes, confirming that 3+ rare earth and 4+ Zr are present across the library, as expected. All Gd $L_3$-edge spectra (Figure 3a) exhibit an intense "white line" at the absorption edge ($E_0$ = 7245 eV) followed by a small peak, a larger peak, and broad oscillations thereafter (not shown). This result is consistent with XAFS studies[59], [60], [61], [62] on fluorite-type structures. Figure 3(b) offers a qualitative comparison, known as fingerprint analysis, of the Gd $L_3$-edge spectra between library position A; bulk powder compositions $(Gd_{0.25}Dy_{0.25}Ho_{0.25}Er_{0.25})_2Zr_2O_7$ (defect-fluorite) and $Gd_2Zr_2O_7$ (pyrochlore); and references $Gd_2Ti_2O_7$ (8-fold coordination), $Gd_2TiO_5$ (7-fold coordination), and $Gd_2O_3$ (6-fold coordination). Reference spectra were obtained from the "Common XAFS materials at BMM" database. $Gd_2Ti_2O_7$ adopts the cubic pyrochlore structure (*Fd-3m*), $Gd_2TiO_5$ adopts an orthorhombic structure (*Pnma*), and $Gd_2O_3$ adopts the cubic bixbyite structure (*Ia-3*). The two peaks following the white line in the library spectrum show similarities to all bulk and reference spectra, however, the position and intensity of the peaks vary. Interestingly, the bulk composition spectra are similar to one another despite $(Gd_{0.25}Dy_{0.25}Ho_{0.25}Er_{0.25})_2Zr_2O_7$ having a defect-fluorite structure type and $Gd_2Zr_2O_7$ having a pyrochlore structure type. This indicates local lattice order/disorder motifs that are similar in the two materials. The small peak in the library spectrum around ~7265 eV is distinct to that in $Gd_2Ti_2O_7$, being slightly shifted to higher energy and mixing more with the ~7285 eV peak. The spectra for $Gd_2TiO_5$ and $Gd_2O_3$ (7-fold and 6-fold coordination) more closely resemble the library spectra. From this qualitative analysis we can say the coordination number of Gd is likely closer to 7, consistent with the long-range defect-fluorite disordering. Additionally, the defect-fluorite structure has been shown to exhibit orthorhombic local ordering[63] and is known to crystalize as cubic or orthorhombic[64] depending on ionic radius size mismatch and synthesis. Considering these results and the resemblance of the library to the $Gd_2TiO_5$ spectra, it is likely we have local orthorhombic-like distortion.

[1] Any mention of commercial products within NIST web pages is for information only; it does not imply recommendation or endorsement by NIST.

The Zr K edge (Figure 3 c,d) exhibits a few characteristic features. First, there is a clear pre-edge shoulder in all library spectra about 6 eV below the absorption edge ($E_0$ = 18008 eV), indicated by a black arrow in Figure 3(d). In general, features in the pre-edge region are due to electronic transitions of core electrons to unoccupied high-energy states near the Fermi level and are sensitive to the local coordination geometry. The first derivative of the Zr K-edge library spectrum highlights this feature as shown by Li et al[58], [65]. This is caused by an electron transition from a lower energy 1s state to a higher energy 4d state. Figure 3(d) compares the Zr K-edge spectra between the library; bulk powder compositions $(Gd_{0.25}Dy_{0.25}Ho_{0.25}Er_{0.25})_2Zr_2O_7$, $Gd_2Zr_2O_7$, and $Ho_2Zr_2O_7$ (defect-fluorite); and cubic (c-$ZrO_2$), tetragonal (t-$ZrO_2$), and monoclinic (m-$ZrO_2$) zirconia polymorphs. These polymorphs provide an excellent reference for the Zr K-edge since they cover 8-fold, distorted 8-fold, and 7-fold coordination geometries for c-$ZrO_2$, t-$ZrO_2$, and m-$ZrO_2$, respectively. The spectra for t-$ZrO_2$ and the library exhibit the pre-edge shoulder indicated by black arrows while it is absent from c-$ZrO_2$,m-$ZrO_2$, and bulk zirconate spectra. This suggests the local environment of Zr in the library is distorted (not ideally cubic). Li et al. observed this shoulder in t-$ZrO_2$ and attributed it to a distortion from centrosymmetry in the Zr-O polyhedron leading toward 4 short and 4 long Zr-O bonds (two distinct tetrahedra). This shoulder becomes weaker as the coordination of Zr approaches 6, as in the pyrochlore structure[58], [61]. Less pronounced splitting of the white line in the library spectrum when compared to the bulk spectra is also consistent with more local atomic disorder being present in the library compositions. In the defect-fluorite structure, cations have an average coordination of 7. However, it is clear that m-$ZrO_2$ does not provide a good match to the library spectrum. Neither is the local order of the library pyrochlore-like, since the coordination environments of the rare earth and Zr cations are not 8 and 6, respectively. The c-

[1] Any mention of commercial products within NIST web pages is for information only; it does not imply recommendation or endorsement by NIST.

$ZrO_2$ and t-$ZrO_2$ spectra exhibit a small peak after the white line around 18040 eV that is absent from m-$ZrO_2$ and the bulk compositions. It is evident from the XANES that the Zr K-edge has features resembling all three zirconia polymorphs. Overall, the XANES spectra imply the local environment of Zr is complex, yet it is consistent across the library/compositions in the study.

Figure 4a)-b) shows the normalized EXAFS FTs for the Gd L3- and Zr K-edges across $(GdDyHoEr)_2Zr_2O_x$ library positions A-H before phase correction as well as bulk powder spectra for the $(Gd_{0.25}Dy_{0.25}Ho_{0.25}Er_{0.25})Zr_2O_7$, $Gd_2Zr_2O_7$ and $Ho_2Zr_2O_7$ (for Zr K spectrum only). There is effectively no difference in the EXAFS across library positions at low R. In the FTs for Gd (Figure 4a), there is a large peak below 2 Å that corresponds to the nearest neighbor (NN) RE-O distance. After the first peak, the Gd L3-edge shows a weak correlation at about 3 Å for the next nearest neighbor (NNN) RE-RE/Zr distance and some weaker correlations around 4 Å, likely resulting from multiple scattering phenomena. The Zr K-edge FT (Figure 4b) features a major peak at 1.6 Å that corresponds to the NN Zr-O distances. Beyond this there is a small peak around 3 Å corresponding to the NNN Zr-Zr/RE distance and surprisingly stronger correlations at higher R that likely result from multiple scattering phenomena. Interestingly, the NNN correlation for Zr is about half the intensity of that for Gd and other rare earths, suggesting the local environment of Zr is more disordered; this is consistent with the signature for the tetragonal distortion in the Zr XANES. All bulk compositions show relatively equal or stronger correlations for neighboring atoms when compared to the library except for the first peak in Gd $L_3$-edge FT for $Gd_2Zr_2O_7$ (corresponding to an average pyrochlore structure). In the Zr K-edge FT for $Gd_2Zr_2O_7$ the NNN shows very strong correlation likely due to the explicit cation ordering in the pyrochlore structure. Overall, the weaker correlations exhibited in the library spectra suggest a higher degree of disordering relative to bulk compositions, particularly for the Zr environment. Figure 4c) provides the RE/Zr-O distances extracted from EXAFS analysis for each cation as a function of library position. In agreement with XANES measurements, the RE/Zr-O distances obtained from EXAFS fitting were very consistent across library positions A-H. As expected, the distances increase with ionic radius of the absorber, ($IR_{Zr} < IR_{Ho} < IR_{Dy} < IR_{Gd}$)[66]. The error bars represent experimental error derived from model fits, the dotted line is the mean distance, and the shaded region represents the mean +/- standard deviation across all library positions. The error bars are larger for spectra generated from smaller k-ranges. MSRDs range from 0.010-0.011 $Å^2$ for Zr-O and from 0.015-0.020 $Å^2$ for RE-O distances. These values reflect the presence of both thermal motion and static disorder. The bulk composition $(GdDyHoEr)_2Zr_2O_7$ consistently exhibits slightly larger cation-oxygen distances that reside outside of the shaded (standard deviation) regions when compared to the library distances, however, these distances fall within respective experimental error bars. Several processing-based variations could explain slightly different local environments in thin film and bulk samples, such as substrate-induced strain effects, differences in the extent of cation homogeneity / short range order, or differences in oxygen vacancy distributions.

[1] Any mention of commercial products within NIST web pages is for information only; it does not imply recommendation or endorsement by NIST.

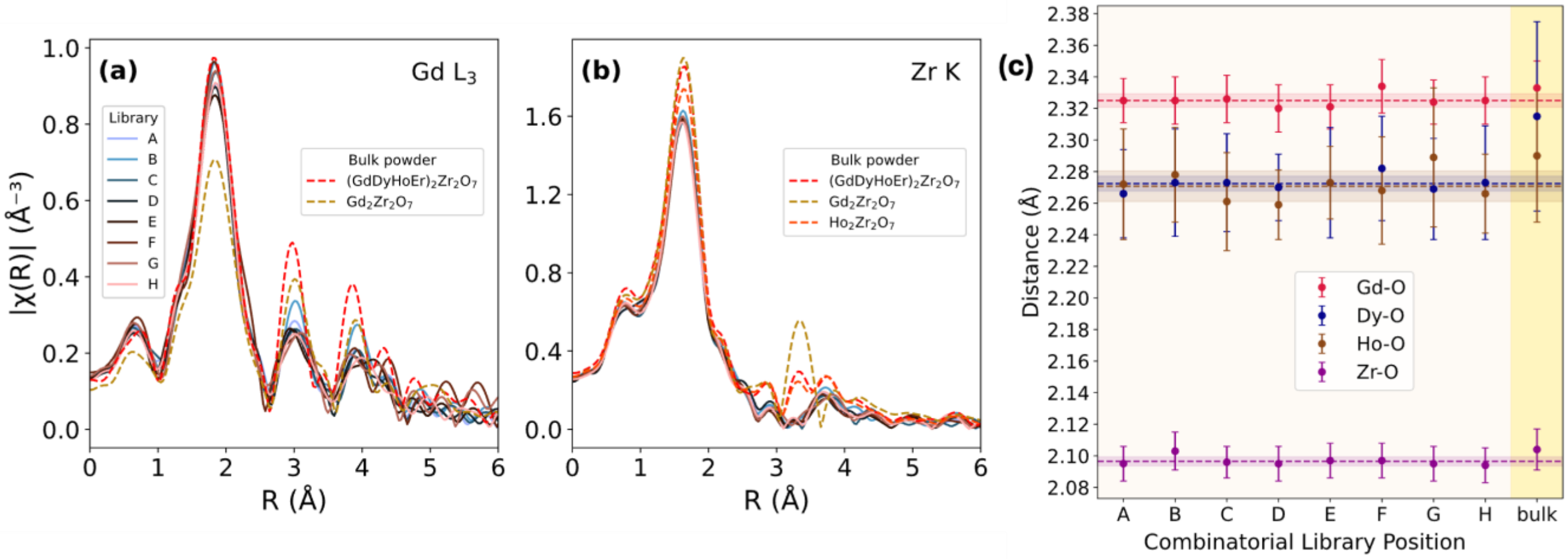

Figure 4: (a) Gd $L_3$-edge Fourier transform (FT) across $(GdDyHoEr)_2Zr_2O_x$ library positions A-H, compared with that of bulk $(Gd_{0.25}Dy_{0.25}Ho_{0.25}Er_{0.25})Zr_2O_7$ (with defect-fluorite average structure) and bulk $Gd_2Zr_2O_7$ (with pyrochlore average structure). (b) Zr K-edge FT across $(GdDyHoEr)_2Zr_2O_x$ library positions A-H compared with that of bulk $(Gd_{0.25}Dy_{0.25}Ho_{0.25}Er_{0.25})Zr_2O_7$ (with defect-fluorite average structure) and bulk $Gd_2Zr_2O_7$ (with pyrochlore average structure), and $Ho_2Zr_2O_7$ (with defect-fluorite average structure). (c) Strip plot showing the spatial distribution of RE/Zr-O distances across the $(GdDyHoEr)_2Zr_2O_x$ library positions A-H and for the bulk composition $(Gd_{0.25}Dy_{0.25}Ho_{0.25}Er_{0.25})_2Zr_2O_7$. Each point includes experimental error bars and the shaded regions represent the mean +/- standard deviation across all library positions (excluding the bulk powder data).

## 3.2 Structural and Topographical Characterization

To determine the crystal structure of the as-deposited $CCRE_2Zr_2O_7$ film, preliminary XRD measurements were performed along a 17-point radial map (see Figure SI1). Interestingly, the probed compositions are crystalline as deposited at room temperature and exhibit a defect fluorite phase (space group $Fm\bar{3}m$)[67] at all points. A comparison of unbiased versus biased XRD patterns in Figure SI4 reveals the zero substrate bias film exhibits a significant 200 preferred orientation.

High-resolution structural analysis of the system was performed via transmission XRD at the Brookhaven National Laboratory NSLS-II. Figure 5 a) shows WPP fit results to 5 XRD patterns, corresponding to the library center and several previously labeled thin film positions. The cubic fluorite model provides an excellent description of the diffraction data, accurately reproducing the positions and widths of the majority of Bragg reflections across the compositional space. Whole-pattern refinements were used to extract the cubic lattice parameter ($a$) and microstrain ($\varepsilon = \Delta d/d$) for each composition, Figure 5 b) and 5 c), providing descriptors of the average lattice dimension and the distribution of lattice spacings, respectively. The refined lattice parameter exhibits systematic variation across the compositional library. Interestingly, while the average ionic radius theoretically peaks nearest the Gd/Zr target location at the bottom center of the library, the measured largest lattice parameter is found to the right and below the film center, between the Gd/Zr and Dy/Zr target positions. This deviation from Vegard-like averaging suggests that the average fluorite lattice is influenced by an additional structural contribution beyond chemical size effects.

---

[1] Any mention of commercial products within NIST web pages is for information only; it does not imply recommendation or endorsement by NIST.

Whole-pattern microstrain, determined from fitting Bragg peak broadening, exhibits modest variation across the library (Figure 5 c), indicating that the overall distribution of lattice spacings changes only gradually with composition. The distribution is a minimum in the same region the maximum lattice dimension is observed. Despite the excellent agreement between calculated and observed diffraction patterns, however, systematic residuals remain for the *h*00 reflections, most prominently the (200) peak (see inset in Figure 5a). Specifically, the location of h00 peaks is found to be shifted to lower 2θ values relative to other families of peaks. . To quantify this behavior, the (200) peak was fit independently and a *d*-spacing displacement was parameterized as

$$\Delta d_{200} = d_{200}^{\mathrm{obs}} - d_{200}^{\mathrm{calc}}$$

where $\Delta d_{200}^{\mathrm{calc}}$ is determined from the refined cubic lattice parameter. The resulting $\Delta d_{200}$ map (Figure 5 d) reveals the reflection-dependent h00 distortion is minimized in the region where the lattice parameter is a maximum and the calculated microstrain is a minimum. The lattice parameter and $\Delta d_{200}$ provide complementary structural descriptors that reveal deviations from the behavior expected for a uniformly expanding cubic fluorite lattice

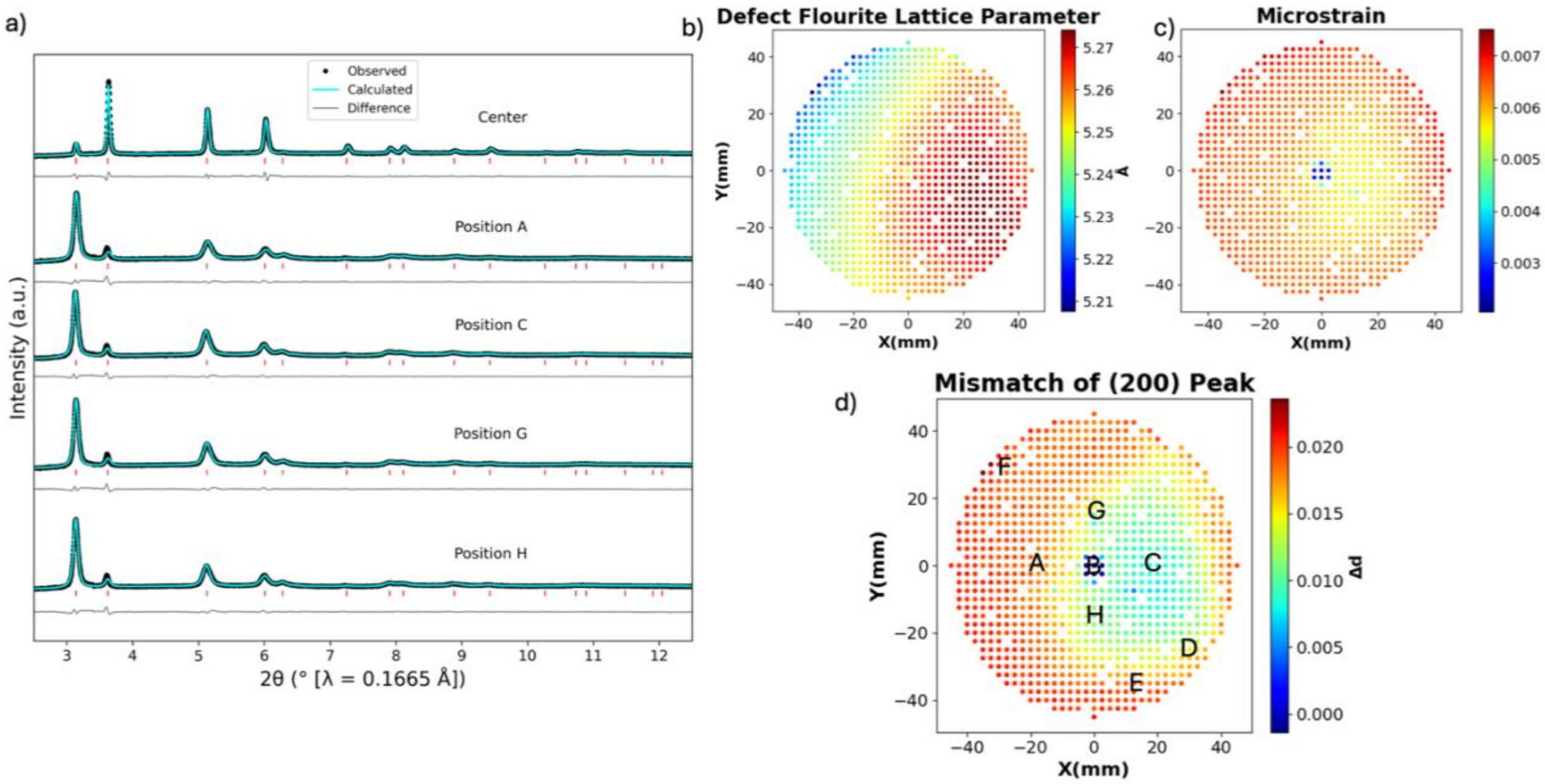

Figure 5: Select results of synchrotron XRD analysis: a) diffraction patterns at center and positions A,C,G, and H (representative of locations within each target quadrant). Calculated whole-pattern fit b) defect fluorite lattice parameter and c) microstrain. d) mismatch of the (200) peak location between whole pattern fit and single peak fit.

Comparing Figure 5 b)-d), an interesting feature is detected at the center of each plot (corresponding to local maxima in lattice parameter, and minima in microstrain and $\Delta d_{200}$), an anomalous ~ 1 cm circle and 3 spots at the vertices of an equilateral triangle with ~ 3 cm edge length. This feature emulates the shape of the substrate sample holder, which has a center pole and a "helicopter-like" sample exchange feature. Normalized diffraction patterns within these regions, relative to the rest of the substrate, reveal strong preferred orientation, consistent with results generated for an unbiased substrate. Thus, it appears that the substrate bias is grounded and suppressed in these regions.

[1] Any mention of commercial products within NIST web pages is for information only; it does not imply recommendation or endorsement by NIST.

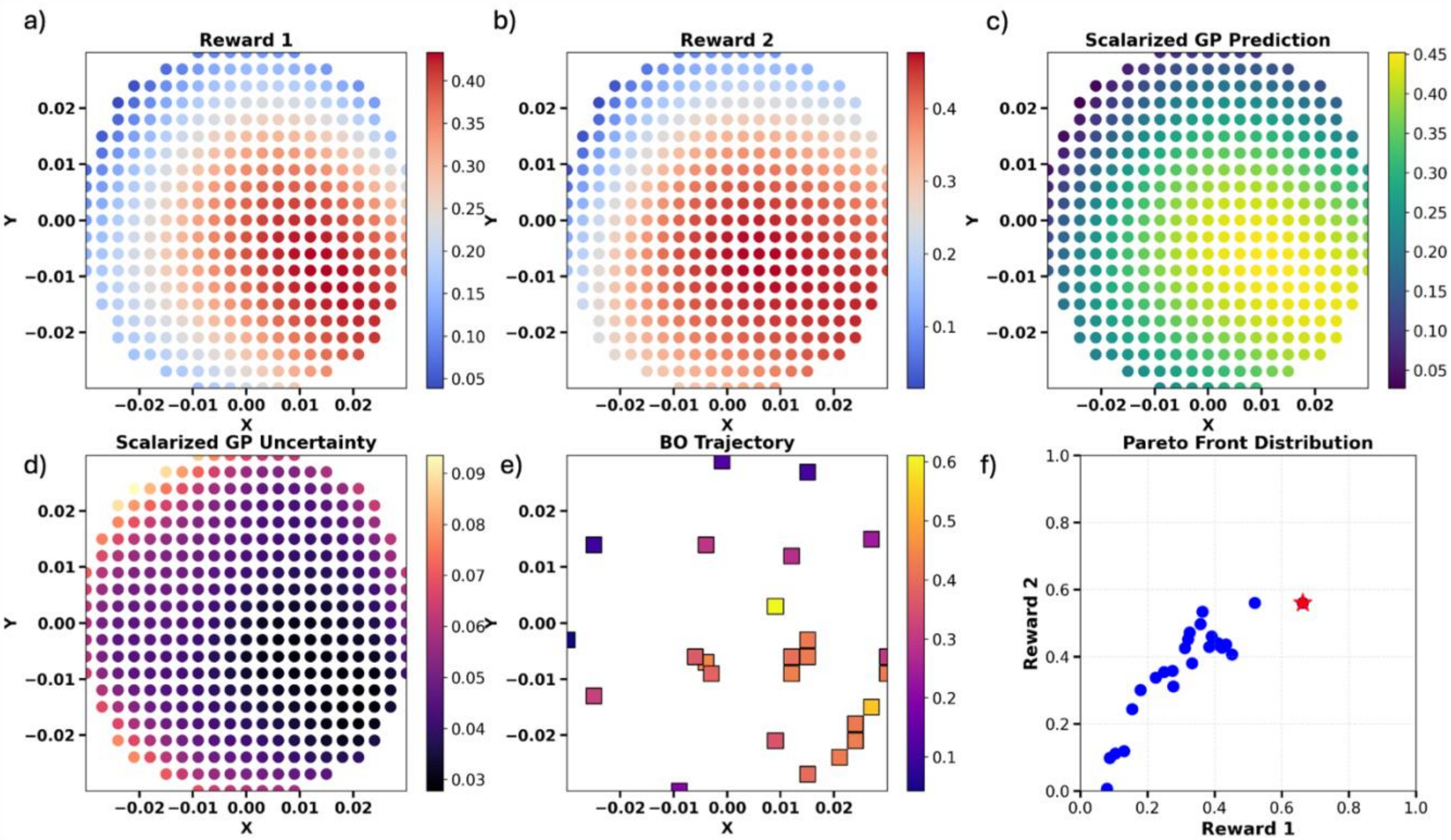

Figure 6: State of the multi-objective Bayesian optimization (MOBO) procedure at iteration 20. a) Spatial distribution of objective 1, corresponding to the particle-size reward, over the feasible design domain. Higher values indicate design conditions associated with more favorable particle-size outcomes under the defined optimization criterion. b) Spatial distribution of objective 2, corresponding to the correlation-based reward. Higher values indicate design conditions yielding stronger correlation performance. c) Posterior mean of the scalarized Gaussian process surrogate used to guide candidate selection, representing the model's current estimate of the scalarized objective across the domain. d) Corresponding posterior uncertainty of the scalarized surrogate, highlighting regions where the model remains less informed. e) Sequence of sampled design points acquired by MOBO up to step 20, illustrating the exploration–exploitation trajectory through the search space. f) Objective-space representation of the sampled solutions, where blue markers denote evaluated designs and the red star indicates the current high-value Pareto-optimal candidate.

From the A-AFM analysis, Figure 5 a) and b) illustrate the resultant reward gradients, representing the mean Voronoi area (reward 1) and correlation length (reward 2). Both parameters exhibit maximum values in the same region between the Gd/Zr and Dy/Zr quadrant of the sample. Figures c)-d) represent the Gaussian process (GP) surrogate and the corresponding uncertainty, showing a high-confidence model. Figures e)-f) correspond to the Bayesian Optimization trajectory and Pareto front distribution of the system, reflecting sparse sampling and high uncertainty in the surrogate models. As additional AFM scans were acquired, the Pareto front expanded and became increasingly dense, revealing a well-resolved trade-off between particle size and structural correlation length. To confirm the MOBO output, a 17-point position absolute AFM scan was performed and resulted in similar trends, Figure SI5. Thus, the AFM analysis concludes that the grain size as determined via AFM is largest in the region that correlates to the largest lattice parameter and lowest (200) plane offset.

[1] Any mention of commercial products within NIST web pages is for information only; it does not imply recommendation or endorsement by NIST.

3.3 Thermal Conductivity Analysis

Fig. 7a) illustrates the thermal conductivity (*k)* map of a 17-point position with linear regression extrapolation to simulate the whole sample space measured via SSTR. A low thermal conductivity region is observed in between the Gd/Zr and Dy/Zr region of the library. A high-resolution thermal conductivity map was also performed on the center 5x5 cm region of the sample with a pixel size of 100 microns and is represented by Fig. 5b). Both measurements report the same low thermal conductivity region of interest between the Gd/Zr and Dy/Zr composition, corresponding to an approximate composition of $(Gd_{0.38}Dy_{0.30}Ho_{0.16}Er_{0.16})_2Zr_2O_x$. From the high-resolution mapping, the region of interest has a *k* range of 0.4 W $(m\ K)^{-1}$ to 0.6 W $(m\ K)^{-1}$.  Notably, this region of interest is not one of equiatomic composition, where most studies fixate (where configurational entropy is maximed)

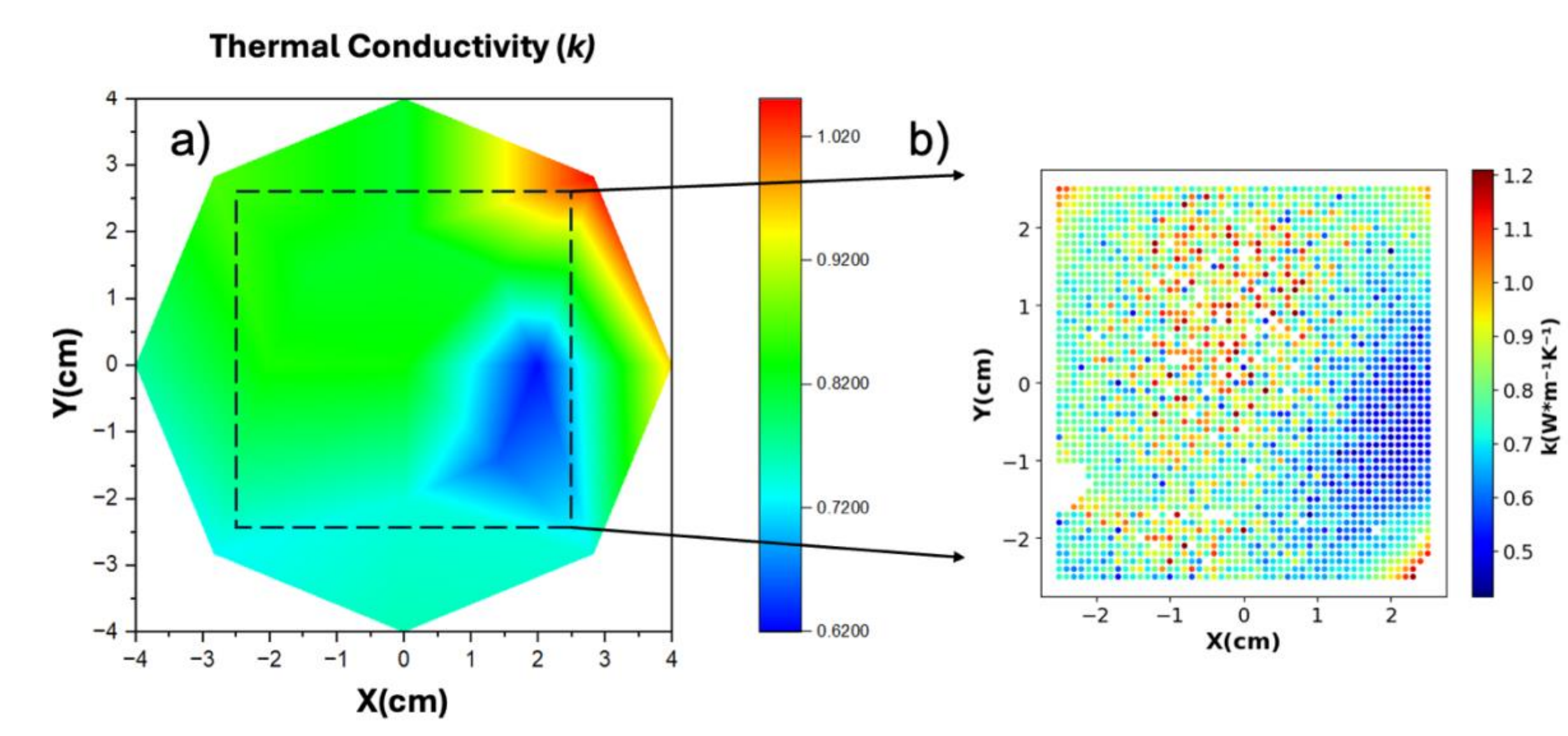

Figure 7: Thermal conductivity heatmaps of a) 17-position and b) 100-micron/pixel 5x5cm CCRE2Zr2O7 thermal conductivity map. The same lower-conductivity region is seen on the right side of each.

4 Discussion

A common framework for understanding thermal transport in compositionally complex fluorite oxides is that increasing cation size disorder enhances phonon scattering, leading to progressively lower thermal conductivity[20]. Under this picture, both the average lattice parameter and thermal conductivity are expected to evolve primarily according to Vegard-like chemical averaging and the magnitude of cation size mismatch. The present results indicate that this description is incomplete in our combinatorial fluorite library.

[1] Any mention of commercial products within NIST web pages is for information only; it does not imply recommendation or endorsement by NIST.

Comparing the thermal conductivity map in Figure 7 to the extracted atomic/microstructure parameters in Figure 5 and the microstructure/topographical information in Figure 6, several trends emerge. The region of lowest thermal conductivity does not coincide with that predicted from cation size disorder alone (Figure 2). Additionally, the films are confirmed nanogranular, with the observed minimum in thermal conductivity corresponding to a maximum in the A-AFM-determined grain size and correlation length (ruling out extrinsic grain size effects). Instead, the region of low thermal conductivity corresponds to compositions exhibiting the largest experimentally measured lattice parameters together with the smallest values of $\Delta d_{200}$. Rather than representing independent observations, the anomalous trends likely reflect a common structural relaxation process within the fluorite lattice. Importantly, XANES and XAFS measurements across this compositional library (Figure 3 and 4) show minimal variation in oxidation state, local coordination, and first-neighbor bonding environments. These results effectively rule out changes in local coordination as the primary origin of the diffraction anomalies and instead indicate that the structural changes responsible for the anomalous lattice expansion and *h*00 behavior occur over longer length scales than those probed by local spectroscopy.

A possible interpretation is that the fluorite lattice accommodates cation-size mismatch through two competing mechanisms. Local weberite-like distortions, previously proposed in rare-earth zirconates and consistent with our XANES and XAFS results, could introduce directional structural distortions that may manifest as reflection-dependent deviations from the average cubic metric, producing finite shifts of $\Delta d_{200}$. [68], [69]Alternatively, the lattice may accommodate this mismatch through a more isotropic structural relaxation that expands the average fluorite lattice while reducing coherent directional distortions. The observed coincidence of anomalous lattice expansion with minimal $\Delta d_{200}$ is consistent with the latter mechanism, suggesting that the lattice becomes more effectively relaxed within a larger average unit cell. Within this framework, cation size disorder remains an important driving force, but it is not by itself the dominant descriptor governing thermal transport. Instead, the critical factor in this compositional library appears to be how the fluorite lattice accommodates that disorder. Understanding how these structural relaxation mechanisms can be controlled through composition and processing, and how they influence thermal transport in both thin-film and bulk fluorite oxides, represents a promising direction for future research.

## 5 Conclusion

A continuous spread combinatorial library of rare earth zirconates (RE=Gd, Ho, Er, Dy) was synthesized by reactive co-sputtering, enabling rapid exploration of composition-structure-property relationships across a four-component fluorite system. Thermoreflectance mapping identified a pronounced minimum in thermal conductivity within the Dy/Gd quadrant, corresponding to an approximate composition of $(Gd_{0.38}Dy_{0.30}Ho_{0.16}Er_{0.16})_2Zr_2O_x$. Importantly, this region does not occur at the equiatomic composition commonly targeted in high entropy ceramics, and neither does it occur where compositional mapping predicts a maximum in cation size disorder. XANES and XAFS show minimal variation in cation valence states, local coordination environments, and nearest-neighbor bond distances across the library, while atomic force microscopy demonstrates the measured grain sizes are nanogranular and that the observed thermal conductivity minima is not due to extrinsic grain size effects. Synchrotron X-ray diffraction further reveals that the lowest thermal conductivity coincides with the largest experimentally measured lattice parameters, despite this region not being predicted by Vegard-like chemical averaging. Whole-pattern refinements indicate only modest variation in microstrain, whereas a systematic reflection-dependent displacement of the $h00$ reflections, quantified by $\Delta d_{200}$, is minimized in the same compositional region. These findings suggest diffraction-derived measures of structural relaxation as promising new descriptors for understanding and optimizing thermal transport in compositionally complex fluorite oxides. Understanding how these relaxation mechanisms can be tuned through composition and processing, and ultimately exploited to tailor thermal transport in both thin-film and bulk fluorite oxides, represents a promising direction for future research. Summarily, we demonstrate a high-throughput method for exploring compositionally complex ceramics via combinatorial thin film co-sputtering and multimodal characterization, uncovering structure-property relationships that would be difficult to identify using conventional discrete-composition studies. This approach provides an efficient pathway for accelerating the discovery and mechanistic understanding of next-generation thermal barrier materials and other complex functional ceramics.

[1] Any mention of commercial products within NIST web pages is for information only; it does not imply recommendation or endorsement by NIST.

Acknowledgements

The authors acknowledge support by the National Science Foundation Materials Research Science and Engineering Center program through the UT Knoxville Center for Advanced Materials and Manufacturing (Grant No. DMR-2309083).

EAS and PEH appreciate support from the Office of Naval Research, Grant Number N00014-25-1-2286. AHJ appreciates support from Air Force Office of Scientific Research under SBIR Contract No. FA9550-25-C-B005.

PDR, DP and JL acknowledge that the SEM and EDS data was conducted as part of a user project at the Center for Nanophase Materials Sciences (CNMS), which is a US Department of Energy, Office of Science User Facility at Oak Ridge National Laboratory

Conflicts of Interest

[1] Any mention of commercial products within NIST web pages is for information only; it does not imply recommendation or endorsement by NIST.

## Supplementary Information

Table SI1: EDS Cation Concentration at nine positions along the combinatorial library

| X(cm) | Y(cm) | Gd at% | Dy at% | Ho at% | Er at% |
|---|---|---|---|---|---|
| -2 | -2 | 44 | 12 | 36 | 12 |
| -2 | 0 | 32 | 14 | 38 | 18 |
| -2 | 2 | 24 | 18 | 34 | 29 |
| -1 | -2 | 50 | 16 | 25 | 11 |
| 0 | 0 | 36 | 21 | 25 | 19 |
| 0 | 2 | 26 | 22 | 25 | 30 |
| 1 | -2 | 52 | 22 | 17 | 10 |
| 2 | 1 | 28 | 30 | 21 | 25 |
| 2 | 0 | 37 | 29 | 18 | 17 |

Table SI2: Simulated Cation Concentrations at nine positions along the combinatorial library

| X(cm) | Y(cm) | Gd at% | Dy at% | Ho at% | Er at% |
|---|---|---|---|---|---|
| -2 | -2 | 45 | 12 | 32 | 11 |
| -2 | 0 | 32 | 14 | 36 | 18 |
| -2 | 2 | 23 | 14 | 36 | 27 |
| -1 | -2 | 48 | 14 | 26 | 12 |
| 0 | 0 | 36 | 20 | 26 | 18 |
| 0 | 2 | 25 | 20 | 25 | 30 |
| 1 | -2 | 48 | 22 | 18 | 12 |
| 2 | 1 | 28 | 32 | 18 | 22 |
| 2 | 0 | 34 | 31 | 17 | 18 |

[1] Any mention of commercial products within NIST web pages is for information only; it does not imply recommendation or endorsement by NIST.

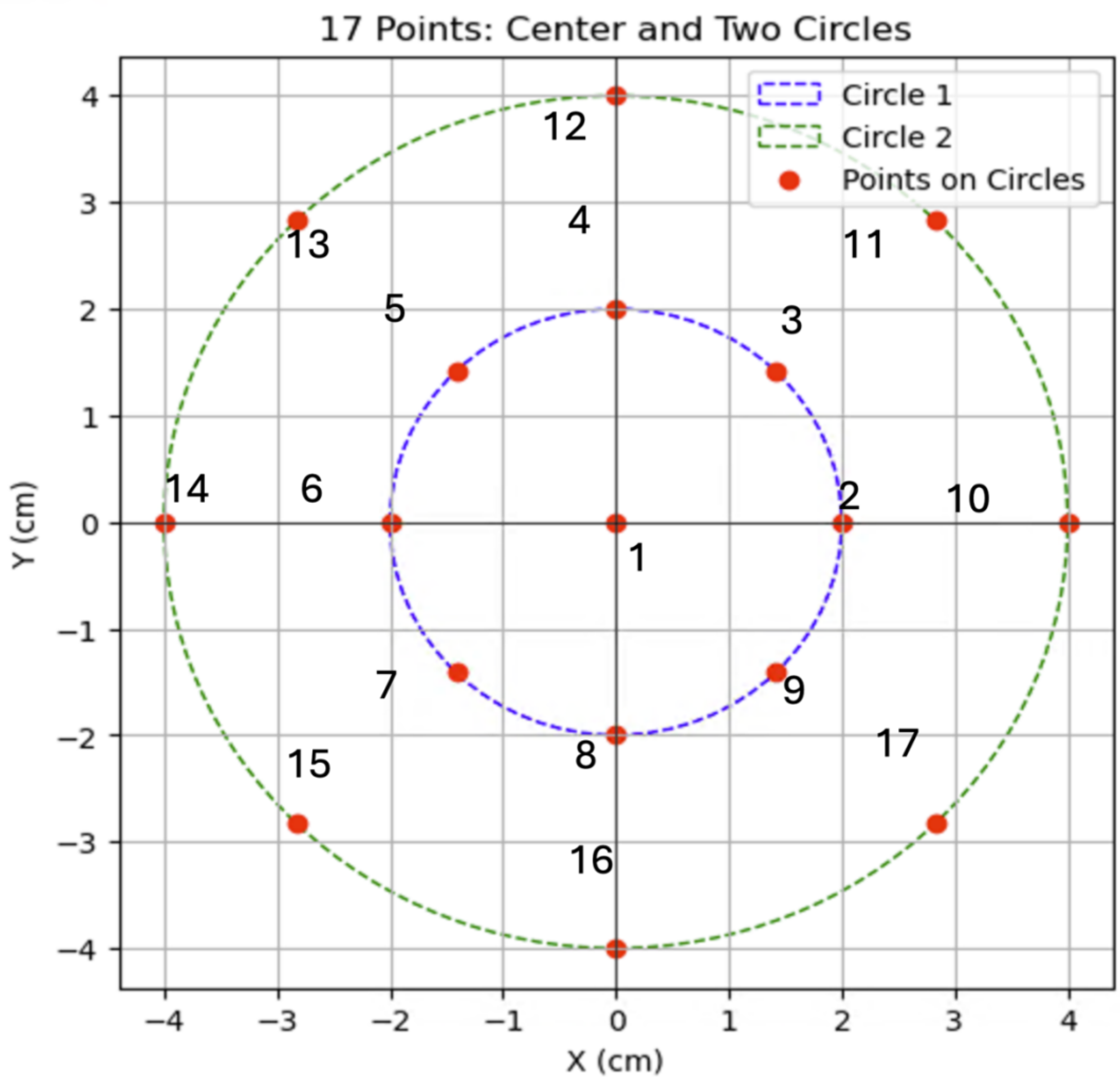

Figure SI1: Locations of 17-positions used for preliminary XRD measurements and ground truth AFM.

[1] Any mention of commercial products within NIST web pages is for information only; it does not imply recommendation or endorsement by NIST.

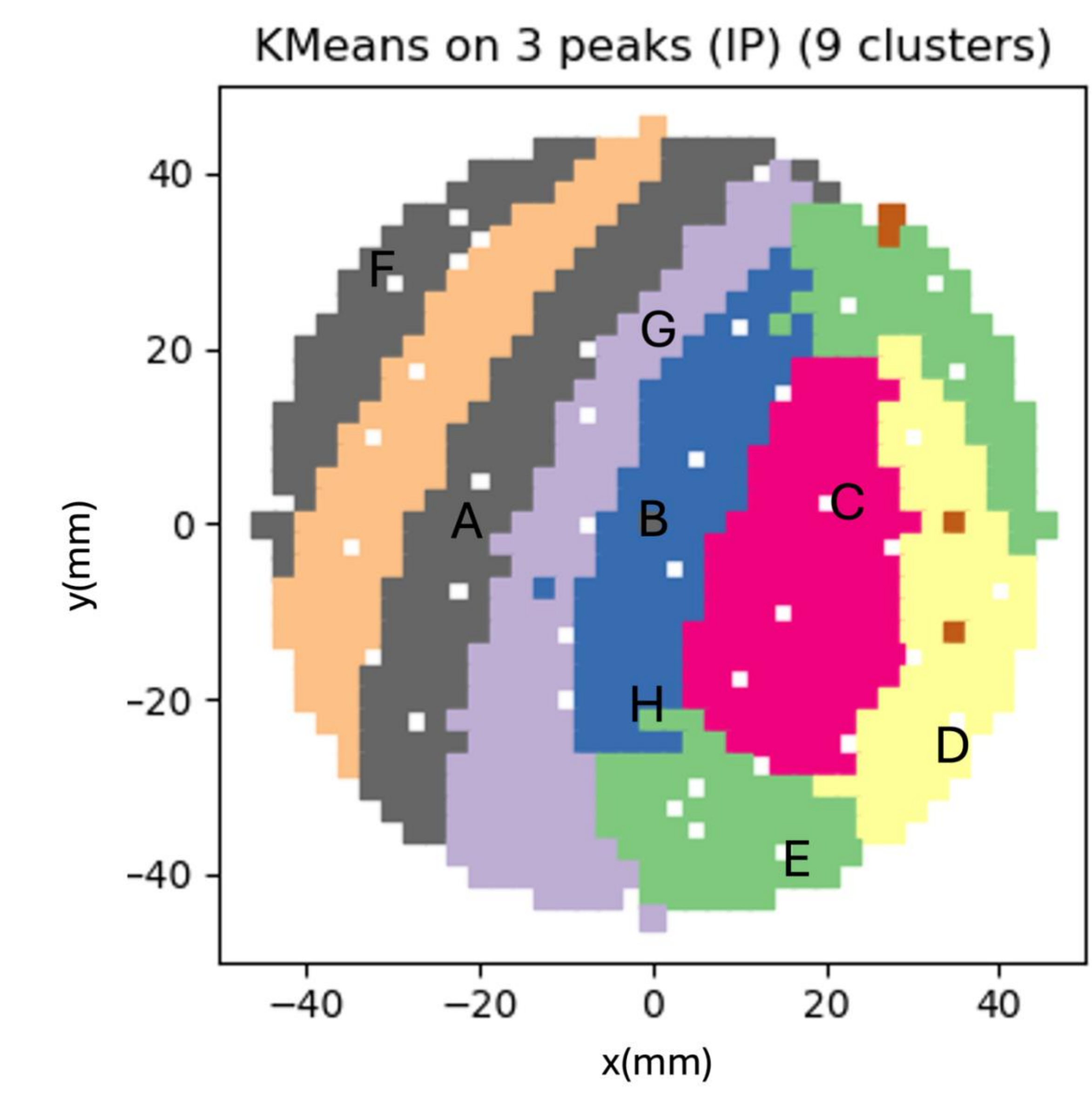

Figure SI2: K-mean cluster map identifying 9 regions of interest

[1] Any mention of commercial products within NIST web pages is for information only; it does not imply recommendation or endorsement by NIST.

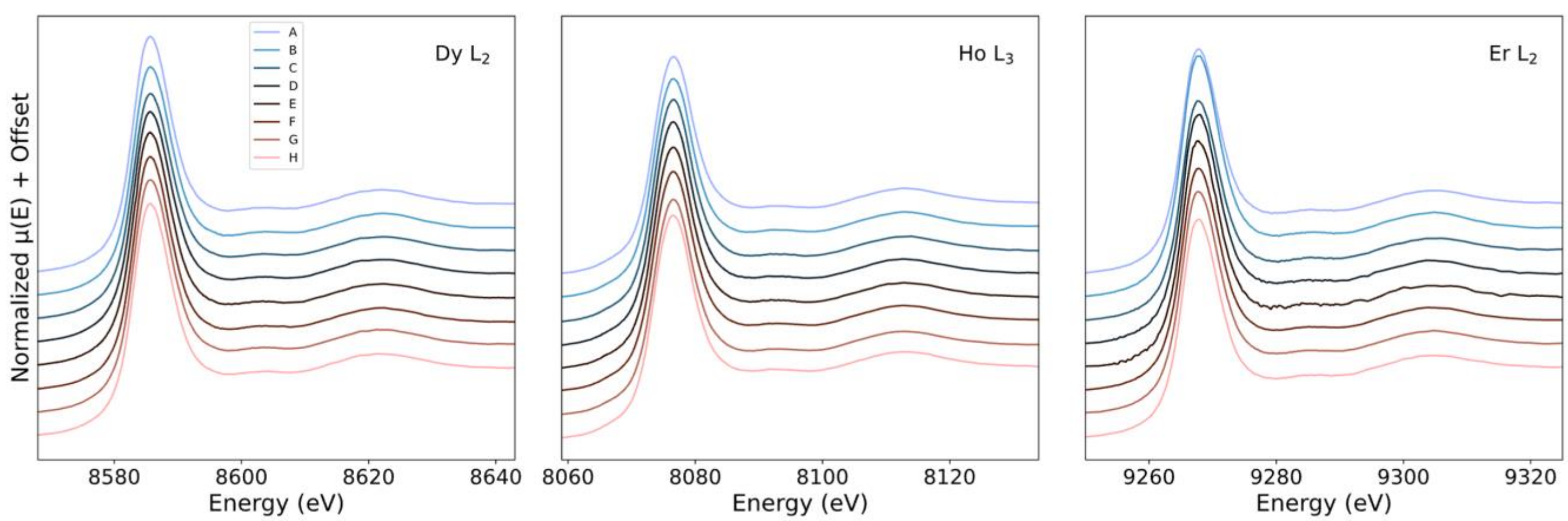

Figure SI3: XANES spectra for rare earth edges Dy $L_2$, Ho $L_3$, and Er $L_2$ with absorption edge energies ($E_0$) of 8583, 8074, and 9265 eV, respectively

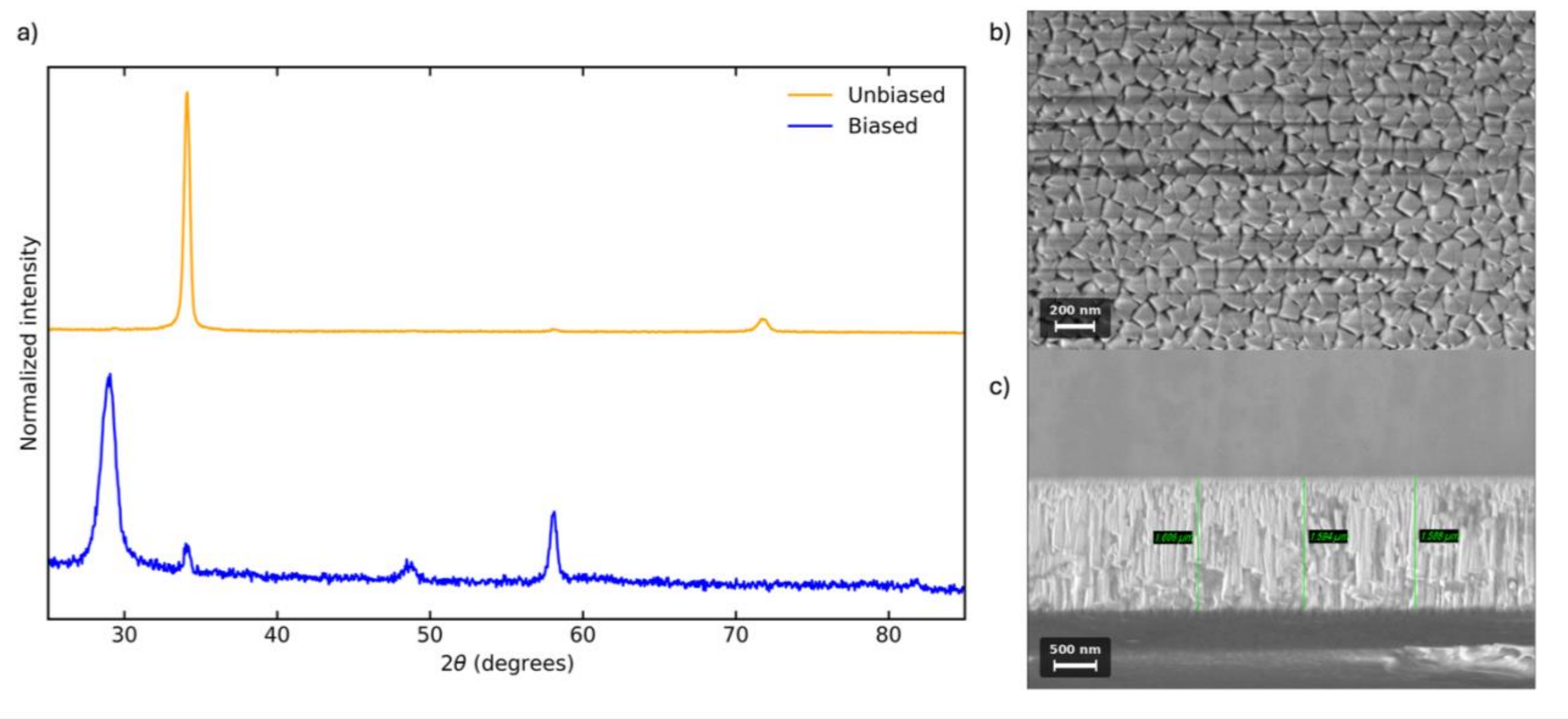

Figure SI4: a) Stacked XRD plot of Unbiased and Biased CCREZrOx thin film. b) SEM image showing porous microstructure of unbiased thin films. c) Cross-sectional SEM image demonstrating porous columnar growth of unbiased thin films.

[1] Any mention of commercial products within NIST web pages is for information only; it does not imply recommendation or endorsement by NIST.

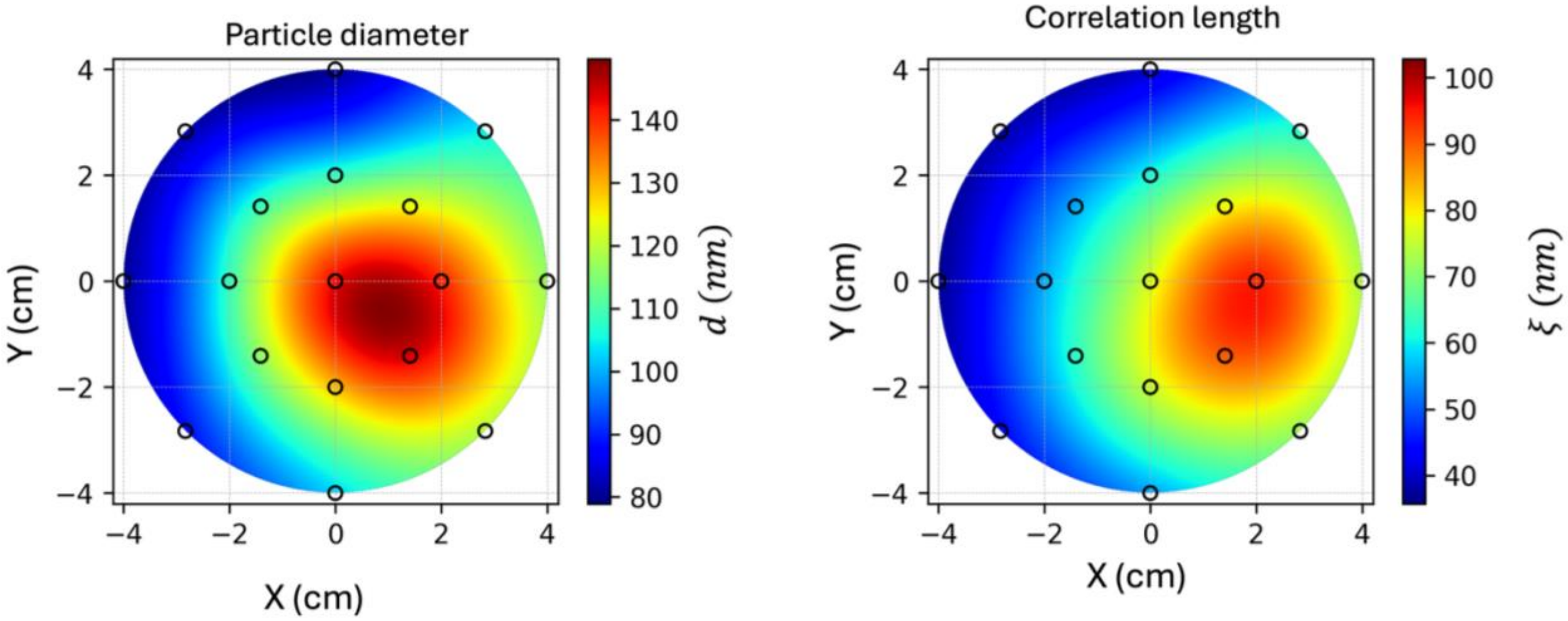

Figure SI5: 17-point spatial distribution maps of the AFM-determined particle diameter and correlation length.

[1] Any mention of commercial products within NIST web pages is for information only; it does not imply recommendation or endorsement by NIST.

---

[1] Any mention of commercial products within NIST web pages is for information only; it does not imply recommendation or endorsement by NIST.

[1] Any mention of commercial products within NIST web pages is for information only; it does not imply recommendation or endorsement by NIST.

[1] Any mention of commercial products within NIST web pages is for information only; it does not imply recommendation or endorsement by NIST.

[1] Any mention of commercial products within NIST web pages is for information only; it does not imply recommendation or endorsement by NIST.